\documentclass[manuscript]{aastex}

\usepackage{rotating}
\usepackage{pdflscape}
\usepackage{mdwlist}

\shorttitle{Likely New Members of the BPMG and ABDMG}
\shortauthors{Schlieder et al.}

\begin{document}

\title{$\beta$ Pictoris and AB Doradus Moving Groups:  Likely New Low-Mass Members}

\author{Joshua E. Schlieder\altaffilmark{1,2}, S\'{e}bastien L\'{e}pine\altaffilmark{3}, and Michal Simon\altaffilmark{1,2}}
\altaffiltext{1}{Department of Physics and Astronomy, Stony Brook University,
    Stony Brook, NY 11794, jschlied@ic.sunysb.edu, michal.simon@stonybrook.edu}
\altaffiltext{2}{Visiting Astronomer, NASA Infrared Telescope Facility (IRTF).}
\altaffiltext{3}{Department of Astrophysics, Division of Physical Sciences, American Museum of Natural History, Central Park West at 79th Street,
   New York, NY 10024, lepine@amnh.org}

\author{Accepted for publication in the Astronomical Journal.}

\begin{abstract}

We present results from our continuing program to identify new, low-mass, members of the nearby young moving groups (NYMGs) using a proper motion selection algorithm and various observational techniques.  We have three goals: 1) To provide high priority targets for exoplanet searches by direct imaging, 2) To complete the census of the membership in the NYMGs down to $\sim$0.1 M$_{\odot}$, and thus 3) Provide a well-characterized sample of nearby (median distances at least twice as close as the Taurus and Ophiuchus SFRÕs), young (8-50 Myr) stars for detailed study of their physical properties and multiplicity.  Our program proceeds as follows:  we apply the selection algorithm to a proper motion catalog where initial selection cuts of candidate members are based on the mean motion of known NYMG members and the proper motions and photometric distances of the candidates.  NYMG membership is investigated further using possible signs of youth, including H$\alpha$ emission and X-ray flux, and then verified through radial velocity (RV) measurements.  We identify TYC 1766-1431-1 (M3), TYC 1208-468-1 and 2 (K3), TYC 7558-655-1 (K5), and PM I04439+3723W and E (M3) as likely members of the $\beta$ Pictoris moving group (BPMG) and TYC 1741-2117-1N and S (K7), TYC 1752-63-1 (K7), TYC 523-573-1 (K7), and TYC 4943-192-1 (M0) as likely members of the AB Doradus moving group (ABDMG).  We also rule out the membership of several BPMG and ABDMG candidates.  To date our program has identified 16 new NYMG members of spectral type K3 or later.
Ê\end{abstract}

\keywords{open clusters and associations: individual ($\beta$ Pictoris moving group, AB Doradus moving group) -- stars: kinematics -- stars: pre-main-sequence}

\section{Introduction}

Young stars exist in the local solar neighborhood in loose associations with each constituent having a common motion through the galaxy, hence the name $\emph{nearby young moving groups}$ (NYMGs).  Rucinski and Krautter (1983) provided evidence that there were young stars far from the nearest star forming regions such as Taurus and Ophiuchus.  Spectroscopic investigations showed that the young star TW Hydrae exhibits classical T-Tauri (TT) properties and later surveys of field stars identified  four more TTs in the vicinity of TW Hydrae (De\ LaReza et al. 1989; Gregorio Hetem et al. 1992).  Kastner et al. (1997) used data from ROSAT (R\"{o}entgen X-ray Satellite) all-sky surveys (Voges et al. 1999, 2000) and high resolution optical spectroscopy of the lithium line at 6708 \AA\ to determine that the 5 stars were indeed a real physical association with an age of $\sim$20 Myr and distance of about 50 pc.  These stars, along with others added later, are known as the TW Hydrae association (TWA).  Several more NYMGs were identified by surveying common proper-motion and distance in the HIPPARCOS catalog of Perryman et al. (1997), identifying sources with far-IR excess emission in the IRAS catalog, and studying stellar kinematics.   Zuckerman et al. (2001) used kinematics and signatures of youth to identify 18 young stars with galactic space motion in common with the A-type star $\beta$ Pictoris.  This association, the $\beta$ Pictoris moving group (BPMG), is spread widely across the sky, and has an estimated age of 12 Myr.  Other proposed NYMGs include the AB Doradus moving group (ABDMG) (Zuckerman et al. 2004, Z04) and the Tucana/Horologium association (Song et al. 2003).  The age of the ABDMG was initially estimated to be $\sim$50 Myr using H$\alpha$ emission strength and color-magnitude diagrams.  More recent studies have shown that the group may be older, 75-150 Myr, and have a common origin with the Pleiades cluster (Luhman et al. 2005; Ortega et al. 2007).  The median distance of stars in the ABDMG is $\sim$35 pc.  The wide distribution of known BPMG and ABDMG members on the sky sets them apart from other NYMGs;  some are accessible from observatories in the northern hemisphere.

With ages between $\sim$10$^7$ and $\sim$10$^8$ years, stars in the NYMGs are the post-T-Tauri stars (PTTS) (Bubar et al. 2007), first proposed by Herbig (1978, p. 171).  PTTS exhibit elevated levels of chromospheric activity, produced by an increased dynamo effect from rapid rotation, and high lithium abundances.  Herbig suggested that PTTS should be numerous but they have proved difficult to identify.  Studying members of NYMGs allows access to this poorly studied period in stellar evolution, which is also the time when gas giant planets are forming and are self-luminous in the IR from gravitational contraction.  The low luminosities of low-mass PTTS produces more favorable contrast ratios between host star and young planet and makes these stars prime targets in searches for sub-solar and planetary companions using high-resolution and high contrast imaging (Lafreni\'{e}re et al. 2008).

Most known members of NYMGs are of spectral type (SpTy) F,G, and K (1.5 - 0.7 $M_{\odot}$), with few of SpTy M ($<$0.7 $M_{\odot}$).  Torres et al. (2006, 2008) identified many new NYMG members in their survey of X-ray sources in the southern hemisphere, but only a small fraction are later than mid-K SpTy.  Since the mass-function of field stars is dominated by low-mass members, peaking at $\sim$0.4 $M_{\odot}$ (Bochanski et al. 2009, p. 977), the census of known members in NYMGs is probably incomplete.  The under-sampling of low-mass NYMG members is a direct consequence of their low luminosity and wide sky distribution.  In this paper we present new results from an ongoing search for new low-mass members of NYMGs.  In \S2 we summarize our application of the proper motion selection technique to the ABDMG and our expansion of the technique to SUPERBLINK proper motion database (L\'{e}pine and Shara 2010 in prep.) to probe candidates down to $\sim$0.1 $M_{\odot}$.  In \S3 we present the NYMG candidate samples and the observational techniques used to identify likely new members.  \S4 describes the likely new members and \S5 summarizes the results and describes our future plans.

\section{Development and Application of the Proper Motion Selection Technique} 

L\'{e}pine and Simon (2009, LS09) introduced a selection technique to identify candidate low-mass members of NYMGs starting with proper motion and photometric data.  Initial selection of candidates is based on the average motion of known NYMG members and the astrometry and photometry of stars in proper motion catalogs.  The mean motion of the group is defined by the U,V, and W space velocities (UVW) (Johnson and Soderblom, 1987) of previously known group members.  The angle that the proper motion of a known member or candidate subtends with the projected mean motion of the group in the local plane of the sky is denoted as $\phi$; $\phi_{max}$ is the maximum value found for the known members of a group (see LS09 eqn. 3).  Our first cut is to select those stars with $\phi$ $\le$ $\phi_{max}$.   Stars are also cut based on their position in an optical/IR color-magnitude diagram (CMD) where absolute K magnitude is determined using the kinematic distance, d$_{kin}$, which is calculated assuming group membership (see LS09 eqn. 7).  A cut in V-K color is applied to select only low-mass stars from the sample, and reduce contamination from background giants.  We refer to stars surviving this round of selection cuts as candidates.  Indicators of youth\ ---\ X-ray emission, H$\alpha$ line emission, and the Li line at 6708\ \AA\ \ ---\  are used to trim the sample further.  We designate stars surviving this cut probable young candidates (PYCs).  PYCs with a high likelihood of group membership are then identified by comparing a predicted radial velocity (RV$_{p}$), calculated assuming group membership, and a measured radial velocity (RV$_{m}$) (See LS09 eqn. 9).  Stars that survive this final selection cut are referred to as likely new members (LNMs) (See Fig. 1 for a pictorial representation of the steps).  LNMs become bona fide new members of an NYMG when parallax data is acquired and the UVW space velocities can be calculated and compared to those of the group. LS09 applied the search technique to a high proper motion ($\mu \ge$ 70 mas yr$^{-1}$) subsample of the TYCHO-2 catalog (H\o g et al. 2000) to search for candidates of the BPMG, identifying 51.  LS09 investigated the membership of 33 of these candidates, finding 4 LNMs.  Here we present a revised search and continue to explore the candidates.  Our search for low-mass BPMG candidates in the $\mu \ge$ 70 mas yr$^{-1}$ TYCHO-2 sample produced 86 candidates, we present follow up of 11.

\subsection{AB Doradus Moving Group Candidates in the TYCHO-2 Catalog}

We have applied our proper motion search in the  $\mu \ge$ 70 mas yr$^{-1}$ subsample of TYCHO-2 stars to identify low-mass ABDMG candidates, adopting ($U_{mg}, V_{mg}, W_{mg}$) = (-8, -27, -14) km s$^{-1}$ as the mean space motion of the group, based on the known member sample of Zuckerman and Song (2004).  In the initial step of the selection technique candidates were restricted to $\phi$ $\le$ 12$^{\circ}$ because the known members of the group also display a range of $\phi$ values, with $\phi_{max_{ABDMG}}$ $\sim$ 12$^{\circ}$.  Our search for ABDMG candidates in the $\mu$ $\ge$ 70 mas yr$^{-1}$ TYCHO-2 sample resulted in 180 low-mass candidates.  This candidate sample is much larger than that of the BPMG when searching in the same TYCHO-2 subsample and is due to the older age of the ABDMG.  Plotting the known members of the ABDMG on CMD shows that they are clearly older than the members of the BPMG.  The absolute magnitudes of ABDMG members of similar SpTy at similar distances are fainter on average (see Fig. 2).  Because of the position of the ABDMG cluster sequence in the CMD the search algorithm is immediately prone to a higher degree of main sequence contamination, selecting older stars whose proper motions align with the mean motion of the NYMG by chance (see Fig. 2).

\subsection{$\beta$ Pictoris Moving Group Candidates in the SUPERBLINK Catalog}

While the TYCHO-2 catalog has been useful in identifying large numbers of candidates of the BPMG and ABDMG, most of the candidates identified are of SpTy's earlier than M3.  This bias is due to the V$\sim$12 limit of the TYCHO-2 catalog.  Using the pre main-sequence isochrones of Siess et al. (2000, S2000), a 10 Myr, 0.4 $M_{\odot}$ star at a distance of 40 pc (the median distance to the BPMG), will have V$\sim$12.  This means that lower mass members of the group generally will not be detected in the TYCHO-2 catalog, and that proper motion catalogs with fainter magnitude limits are needed to detect the members with later SpTy.

The SUPERBLINK (SBK) proper motion database is a high precision astrometric catalog generated by digitally superposing and then blinking images from different epochs of the Digitized Sky Surveys to measure proper motions (L\'{e}pine and Shara 2005, 2010 in prep.).  SBK is statistically complete to V$=$19.0, and has a faint magnitude limit of V$=$21.0, allowing access to BPMG candidates with masses down to $\sim$0.1 $M_{\odot}$.  The bright magnitude limit of the catalog is V$\sim$10.0, complementing the magnitude range covered in TYCHO-2.  The catalog contains 1.7 million stars above the celestial equator with proper motions $\ge$ 40 mas yr$^{-1}$  and proper motion errors $\leq$ 8 mas yr$^{-1}$.  While $\phi_{max_{BPMG}}$$\sim$25$^{\circ}$ we have restricted our search to include only those candidates with $\phi$ $\le$ 12.0$^{\circ}$, in order to reduce interlopers from the extremely large initial sample; there are only 2 known members with larger $\phi$ which are outliers in the more restricted case.  Our search of the SBK catalog for low-mass BPMG candidates returned 195 stars (See Fig. 3).  The SBK catalog BPMG search returned more candidates than the search in the TYCHO-2 subsample even though SBK only covers the northern hemisphere.  This is because the 40 mas yr$^{-1}$ proper motion limit allows us to probe a larger volume, and because the lower magnitude limit makes it significantly more sensitive to stars with lower masses.

\section{Identification of Likely New Members Among the Candidates}

\subsection{Candidate Samples}

Table 1 lists the 19 BPMG candidates targeted for follow up in this analysis.  For each candidate the columns list the catalog name, in either the TYCHO-2 or SBK catalogs, the Hipparcos catalog number if available, the ICRS epoch J2000 coordinates, the proper motions from either the TYCHO-2 or SBK catalog, the catalog V magnitude, the K magnitude from the Two-Micron All Sky Survey (2MASS) catalog of point sources (Skrutskie et al. 2006), the kinematic distance (d$_{kin}$) (see LS09 Eqn. 7), the parallactic distance (d$_{\pi}$) if available, the ROSAT X-ray count rate and hardness ratios if available, the H$\alpha$ equivalent width (EW) if detected, and some notes on the membership of the candidate.  There is some cross-contamination between BPMG and ABDMG candidate samples because the groups have very similar U and W space velocities.  Table 2 lists the 41 TYCHO-2 ABDMG candidates targeted for follow up in the same format as Table 1.  

\subsection{Trigonometric Distances}

If a candidate actually belongs to an NYMG, then its kinematic distance, $d_{kin}$ (i.e. the distance calculated assuming group membership), should be consistent with the distance measured by trigonometric parallax, d$_{\pi}$ (see LS09).  Of the 19 BPMG candidates investigated here, only one, TYC 1766-1431-1 (HIP 11152), has a trigonometric parallax, measured by Hipparcos.  Its kinematic and parallactic distances match very well and it is targeted for further follow up.  In the ABDMG candidate sample nine stars are Hipparcos stars.  One ABDMG candidate has matching d$_{kin}$ and d$_{\pi}$ values,   TYC-8513-952-3 (HIP 22738).  The remaining eight have inconsistent values of $d_{kin}$:  five of them have d$_{kin}$ overestimating their true distances and are likely background objects moving with greater transverse velocities than the ABDMG, the remaining three have d$_{kin}$ $<$ d$_{\pi}$ and must be foreground objects moving slower than the group.  After ruling out these candidates as possible ABDMG members we focus our follow up observations on the 19 BPMG candidates and the remaining 33 ABDMG candidates.  
 
\subsection{Chromospheric Activity: X-ray Emission}

Following LS09 we survey the remaining NYMG candidates for counterparts in both the ROSAT All-Sky Bright Source catalog (RASS-BSC) and the ROSAT All-Sky Faint Source catalog using 50$^{\prime\prime}$ as the largest accepted positional offset between candidate and X-ray source.  In the BPMG candidate sample six stars have ROSAT counterparts,  including five stars drawn from the TYCHO-2 sample and one drawn from SUPERBLINK.   Nine ABDMG candidates show positional coincidence with X-ray emitters in the ROSAT catalogs.  
	
Of the six BPMG candidates with counterparts in the ROSAT catalogs, five have positions placing them less than 15$^{\prime\prime}$ from the X-ray source.  TYC 1766-1431-1 (HIP 11152) has a larger positional offset.  The proposed counterpart to TYC 1766-1431-1 is 23.5$^{\prime\prime}$ from the star with a positional uncertainty of 17$^{\prime\prime}$.  Inspection of the DSS and 2MASS images shows that there are several optical and IR sources near the star, but none are as close to the X-ray source position as TYC 1766-1431-1, even though it is outside of the quoted positional error.  We therefore target TYC 1766-1431-1 for RV measurements.  All 9 of the ABDMG candidates are within 15$^{\prime\prime}$ of the ROSAT X-ray source. 

\subsection{Chromospheric Activity: Atomic Hydrogen Emission}

Because some known NYMG members do not have ROSAT X-ray counterparts (LS09), a lack of X-ray emission is not sufficient to rule out a star as an NYMG candidate and another indicator of youth must be pursued.  Hydrogen line emission may also be used as a diagnostic for youth.  We observed a subsample of the low-mass NYMG candidates presented here to identify those with detectable H$\alpha$ emission using the 1.5 meter telescope operated by the SMARTS consortium at the Cerro Tololo Inter-American Observatory  (CTIO).  We used the RC Spectrograph to obtain low resolution spectra using the same instrumental setup described in LS09.  These data were obtained by service observers and processed by F. Walter following the procedures described in Walter et al. (2004). 

Seventeen BPMG candidates and 37 ABDMG candidates were observed at SMARTS.  Nine NYMG candidates observed exhibit H$\alpha$ in emission, 3 in the BPMG sample and 6 in the ABDMG sample.  The equivalent widths of the H$\alpha$ lines in these stars were estimated using software developed in the IDL programming language; the measurements are presented in Tables 1 and 2.  As it turns out,  all of the stars where H$\alpha$ was detected also have X-ray counterparts in ROSAT.  Conversely, all the candidates with X-ray counterparts were found to have H$\alpha$ in emission,  except TYC-7558-655-1, for which H$\alpha$ remained undetected (n.d.).  The remaining 45 candidates with undetected H$\alpha$ and no X-ray counterpart are ruled out as possible members of the BPMG or ABDMG, leaving 14 candidates for RV follow up (\S 3.5).

\subsection{Radial Velocity Confirmation}

All radial velocity (RV) measurements were obtained using the Cryogenic Echelle Spectrometer (CSHELL; Greene et al., 1993) at the NASA Infrared Telescope Facility (IRTF) 3-meter telescope using the same instrumental setup as in LS09.  The spectra were taken as a series of beam switched 300s integrations for targets with H $\le$ 7.0 and 600s integrations for targets with H $>$ 7.0 and were extracted using the procedures described in Bender and Simon (2008).  RVs were measured by cross-correlation with a sample of low-mass RV standard spectra described in Mazeh et al. (2002) and Prato et al. (2002).  These spectra were rotationally broadened to generate a suite of templates at different $v$sin$i$'s which can be used to estimate the projected rotational velocity of the target star.  The $v$sin$i$ increment of the templates is 2 km s$^{-1}$, which sets the uncertainty in our measured values. We used cross correlation software kindly provided by Chad Bender (priv. comm.) which is capable of measuring velocities of unresolved companions.  Stars in our PYC sample may be spectroscopic binaries and introduce contamination, thus, we test the young candidates for which we obtain RV measurements as double lined systems.  No double lined spectroscopic binaries were identified in the PYCs that we observed.
 
To assess the scatter expected between the predicted radial velocities (RV$_{p}$), and the measured values (RV$_{m}$) we measured the RVs and projected rotational velocities for 9 known ABDMG members.  Table 3 shows that all have predicted values within 3.0 km s$^{-1}$ of the measurements.  Similarly LS09 observed 6 known members of the BPMG, finding the largest difference between RV$_{p}$ and RV$_{m}$ ($\Delta$RV) to be 4.5 km s$^{-1}$.  We will therefore accept $\Delta$RV $\le$ 5 km s$^{-1}$ as the criterion for designating a probable young candidate (PYC) a likely new member (LNM).  We note that the predicted radial velocity is calculated assuming that the candidate has the mean UVW space velocity of the group, which introduces an uncertainty in RV$_{p}$ comparable to the scatter in UVW velocities of known group members.  To allow for this uncertainty and the uncertainty in the measured RV we chose a conservative cut off for $\Delta$RV which may introduce contaminants to the final sample but leaves less likelihood of rejecting potential bona fide members.   The SpTys in Table 3-4 were determined by visual comparison with the standards, and have uncertainties of $\pm$1 spectral subclass. 

 Of the eight PYCs remaining in the BPMG candidate sample we measured RVs for three, including two from the TYCHO-2 sample and one from the SBK sample.  Table 4 lists stars for which we have obtained RV measurements plus stars for which accurate RVs were found in the literature.  The table is organized into columns designating the star's catalog name, alternate name, IRCS epoch J2000 coordinates, predicted RV, measured RV, projected rotational velocity, d$_{kin}$, SpTy, and membership status.   Two of the candidates observed, TYC 1766-1431-1 (HIP 11152) and PM I04439+3723W, have RV$_{m}$ within 4.4 km s$^{-1}$ of RV$_{p}$, comparable to the observed differences in known group members, and are identified as LNMs of the BPMG.  The star PM I04439+3723W has a faint common proper motion companion (named PM I04439+3723E) listed in the SUPERBLINK catalog with an angular separation of $\sim$9$^{\prime\prime}$.  Since the two stars appear to form a bound system, we identify PM I04439+3723E as a LNM of the BPMG as well.  The star TYC 1158-1185-1N is found to have $\mid$RV$_{p}$ - RV$_{m}$$\mid$ = 7.5 km s$^{-1}$, and is thus not retained as a LNM of the BPMG.  TYC 1158-1185-1N is however found to be a visual binary with the secondary (TYC 1158-1185-1S) only $\sim$4$^{\prime\prime}$ away, noticed while acquiring the target for RV observations.  We measured the RVs of both stars, which differ by only $\sim$0.1 km s$^{-1}$, indicating they are probably a bound pair.  TYC 1186-706-1, reported to be a LNM of the BPMG in LS09, was re-observed using CSHELL.  Direct imaging data also revealed the star to be a visual binary, with an angular separation of $\sim$2$^{\prime\prime}$.  We only obtained RV data for the primary but we consider the newly identified secondary to be a LNM of the BPMG also.    
  
 The 2 remaining BPMG PYCs, TYC 1208-468-1 and TYC-7558-655-1, were not observed by us but have RVs published in the literature.  The RV$_{m}$ of these stars are noted with superscripts in Table 4.  The RV of TYC 1208-468-1, a known visual binary, was reported by Jeffries (1995) and matches well with our predicted value.   We accept both components of this system as LNM members of the BPMG.  Torres et al. (2006) (T06) measured the RV of TYC-7558-655-1; the RV was also reported by Zwitter et al. (2008) (Z08) in their Radial Velocity (RAVE) survey second data release.  The measured values are 16.3$\pm$1.1 km s$^{-1}$ and 14.5$\pm$2.4 km s$^{-1}$ respectively.  Both of these values match very well with our RV$_{p}$.  TYC-7558-655-1 is a LNM of the BPMG using our membership criteria but we note that it does not satisfy T06's criteria (see \S 4.1 below).    
  
In the ABDMG PYC sample, five of the eight stars were observed using CSHELL (Table 4).  We identify TYC 1741-2117-1N and S, TYC 1752-63-1, TYC 5899-26-1  and TYC 523-573-1 as LNM of the ABDMG with $\mid$RV$_{p}$ - RV$_{m}$$\mid$  $\le$ 4.1 km s$^{-1}$ in each case.  TYC 1741-2117-1N and S are a visual binary system with a separation of $\sim$2$^{\prime\prime}$ estimated from CSHELL direct imaging data.  The similar radial velocities of the components suggest that they are a bound system.  Our measurements rule out TYC 7002-2219-1 as a LNM of the ABDMG, showing a difference in the predicted and measured RV of $\sim$15 km s$^{-1}$.  

The 3 remaining PYCs in the ABDMG candidate sample have RV measurements available in the literature.  The RVs of TYC 8513-952-3 (HIP 22738) and TYC 9337-2134-1 were measured by T06.  Torres et al. (2008) identify TYC-8513-952-3 (HIP 22738) and TYC 5899-26-1 as new members of the ABDMG.  The RVs reported for the 2 PYCs match well with our measured (in the case of TYC 5899-26-1) and predicted velocities and we reconfirm their membership in the ABDMG.  TYC 8513-952-3 (HIP 22738) is a visual binary with a separation of $\sim$8$^{\prime\prime}$ (T06).  The radial velocity of TYC 4943-192-1 is reported in Z08 as 1.4$\pm$4.6 km s$^{-1}$, RV$_{m}$ is in excellent agreement with RV$_{p}$ and we identify the star as a LNM of the ABDMG.  We will follow up with our own RV measurements to improve the uncertainty of the measurement.  The reported RV rules out TYC-9337-2134-1 as a LNM of the ABDMG.

\section{Likely New Members}

\subsection{$\beta$ Pictoris Moving Group}

{\it TYC 1766-1431-1}:  We identify TYC 1766-1431-1 (HIP 11152) as an M3Ve dwarf at a predicted distance of 29.4 pc.  The Hipparcos distance is 30.8$\pm$3.3 pc, matching very well with our predicted distance.  The star was first identified as a high proper motion M type star (LP 353-51) by Luyten (1979) from photometric observations at the Palomar Observatory.    Zickgraf et al. (2003) (Z03) identified TYC 1766-1431-1 as the optical counterpart to X-ray source 1RXS J022326.7+224344.  

{\it TYC 1208-468-1 and 2}:  We identify TYC 1208-468-1 and 2 (BD+17 232AB) as a young, active, late-type system at a predicted distance of 52.6 pc.  The separation of the components is 1.8$^{\prime\prime}$ (Couteau 1970).  Jeffries (1995) detected H$\alpha$ and Li and measured $v$sin$i$ $>$ 10 km s$^{-1}$ in both components.  Jeffries lists the primary as SpTy K3Ve and estimates a distance of 63$\pm$20 pc. 

{\it TYC 7558-655-1}:  TYC 7558-655-1 (CD-44 753) is a young, late-type star at a predicted distance of 35.7 pc.  T06 identify the star as a K5Ve pre-main sequence star and measure a 50 m\AA$\ $lithium EW.  The lithium EW is small compared to that of known BPMG members of similar SpT (See T08, Figure 9) and may be the reason T06 reject the star as a possible member of the BPMG.  We include it as a LNM because it satisfies the criteria we apply.

{\it PM I04439+3723W and E}:  PM I04439+3723W is an M3Ve dwarf, at a predicted distance of 76.9 pc.  The short rotational period, 4.3 d (Norton et al. 2007, N07), may account for the relatively large rotational velocity, $v$sin$i$ $\sim$ 8 km s$^{-1}$.  The star has a common proper motion companion (PM I04439+3723E) listed in the SUPERBLINK catalog which has an angular separation of $\sim$9$^{\prime\prime}$, the position of the star is listed in Table 4.  The companion is also listed in the NOMAD catalog (Zacharias et al. 2004, p.1418).  The presumed secondary is quite red, with V - K = 5.26, S2000 PMS evolutionary models predict an M5 SpTy assuming the age of the BPMG (12 Myr).  If we accept the predicted distance of PM I04439+3723W to be the true distance to the system the companion lies at a projected separation of $\sim$540 AU.  The companion should also be considered a likely member of the BPMG.

\subsection{AB Doradus Moving Group}

{\it TYC 1741-2117-1N and S}:  TYC 1741-2117-1NS (BPM 84322) is a $\sim$2$^{\prime\prime}$ visual binary with K7Ve/K7Ve components at a predicted distance of 47.6 pc.  At this distance the projected separation of the system is 95 AU.  Stephenson (1986a) identified the system (assuming the stars were unresolved) as a K5 dwarf with a high proper motion, consistent with our characterization.  N07 identify the system as a photometric variable with a period of 3.1555 days, which is consistent with the relatively large $v$sin$i$'s of the components (see Table 4).  Z03 identified TYC-1741-2117-1NS as the optical counterpart to X-ray source 1RXS J003408.7+252342.

{\it TYC 1752-63-1}:  TYC-1752-63-1 (StKM 1-174), which we identify as a K7Ve dwarf, lies at a predicted distance of 38.5 pc.  Stephenson (1986b) first identified the star as a red dwarf based on the strength of its sodium D line, classifying it as SpTy K5.  This subtype is broadly consistent with our own CCD classification.  The star was also identified by Z03 as the optical counterpart to X-ray source 1RXS J013723.4+265709 in the RASS-BSC.  N07 identified the star as a variable with variation at the 5$\%$-6$\%$ level in the optical regime and a period of 1.0852 d.

{\it TYC 523-573-1}:  TYC 523-573-1 (BD+05 4576), which we identify as a K7Ve dwarf, is at a predicted kinematic distance of 38.5 pc.  The star also appears in the Stephenson (1986a) catalog of high proper motion dwarf K and M stars.  Stephenson classifies the star as a K type from from visual inspection of the plate spectrogram.

{\it TYC 4943-192-1}:  We identify  TYC 4943-192-1 as a young, late type star  and estimate its distance to be 45.5 pc.  Z08 observed the star measuring a radial velocity of 1.4$\pm$4.6 km s$^{-1}$, matching well with our RV$_{p}$.  The star is identified as RAVE J121518.4-023728 in the RAVE second data release.  

\subsection{Comparison to Moving Group Cluster Sequences}

The LNMs we have identified survived our selection cuts but interlopers may still remain in the sample and comparison to group cluster sequences is useful.  We have consolidated all LNM systems identified thus far and compared them with known group members in color-magnitude diagrams (Fig. 5).  The absolute K magnitude has been calculated using d$_{kin}$ for both known members and LNMs.  Figure 5 shows that our LNMs are consistent with the known members of the BPMG and ABDMG.  We are pursuing parallax and lithium measurements for the LNMs as a definitive check of NYMG membership. 

\section{Summary}  
  
 Using the proper motion selection technique described by LS09 we identify six likely new low-mass members of the BPMG and five likely new low-mass members of the ABDMG\footnote{Table 4 lists 8 PYCs as LNMs of the ABDMG, 3 of these stars were identified as new members by Torres et al. 2008 while this work was in progress.}.  Definitive assignment of membership awaits parallax and lithium measurements to confirm the presently derived kinematic distances (and thus UVW space velocities) and suspected youth.  Of the 11 LNMs identified by us in this work, 6 stars are components of visual binaries, 2 systems in the BPMG and 1 in the ABDMG.  Together with the results in LS09, the proper motion selection technique has now yielded at least 16 stars as likely members of these groups.  Many of the likely members identified are in the northern hemisphere, the number of probable BPMG members in the north has doubled as a result of this work (See Fig. 4).  All the primaries are of SpTy later than K2 but the latest spectral type represented is M3.  The fact that no LNMs later than SpTy M3 have been identified is probably a consequence of the V$\sim$12 mag. limit of the TYCHO-2 catalog, which has supplied most of the candidates investigated thus far.  In the present work we describe our first use of the SUPERBLINK database which is complete to V$\sim$19 mag..  We are continuing our search for SUPERBLINK candidates in the BPMG and ABDMG and will apply our search procedure to the TW Hydrae and Tucana/Horologium NYMGs; it is our hope that LNMs of SpTy later than M3 will be discovered in the search.
 
 The crucial first step in our search procedure for a given NYMG identifies candidates according to the angle that the proper motion of the candidate subtends with the projected mean motion of the group in the local plane of the sky ($\phi$).  The cosine of $\phi$ is the scalar product of these 2 vectors.  As likely new members of an NYMG are identified, for example by T08 and our work, the NYMG's mean motion, and the intrinsic spread in the known member cluster sequence, becomes better determined.  Accordingly, we will be able to refine our candidate searches in any proper motion catalog using revised NYMG known member samples.  The continued application of our search technique to the astrometric catalogs discussed in this work and other proper motion catalogs such as the PPM-Extended catalog (PPMX) (R\"{o}ser et al. 2008) and the LSPM-South catalog (L\'{e}pine priv. comm.) using revised known member samples is the focus of current work and will be presented in forthcoming publications.  With future applications of the search procedure we look forward to the identification of more high priority targets for direct imaging exoplanet searches as well as progressing towards a complete census of the members of the NYMGs; providing a well-characterized sample of nearby, young, low-mass stars for study of their properties, both physical and kinematic.

 \
\
\
\  
\acknowledgments
We thank the referee for a prompt and helpful report.  We thank C. Bender for sharing with us his software for the extraction of CSHELL spectra and for their analysis by correlation techniques.  We also thank R.W. White for discussing his Li line measurements in advance of publication and H. Harris for including some of the stars discussed here in his parallax program.  The work of J.E.S and M.S. was supported in part by NSF grants AST 06-07612 and 09-07745.  The work of S.L. was supported by NSF grants AST 06-07757 and AST 09-08406. This publication makes use of data products from the Two Micron All Sky Survey, which is a joint project of the University of Massachusetts and the Infrared Processing and Analysis Center/California Institute of Technology, funded by the National Aeronautics and Space Administration and the National Science Foundation.  This research has made use of the SIMBAD database, Aladin, and Vizier, operated at CDS, Strasbourg, France.

\clearpage

\begin{sidewaystable}
\scriptsize
\hspace{-2.5cm}
\begin{tabular}{llrrrrrrrrrrrrr}
\multicolumn{15}{c}{\textbf{Table 1}}\\
\multicolumn{15}{c}{\textbf{Candidate $\beta$ Pictoris Moving Group Members}}\\
\hline
\multicolumn{15}{c}{TYCHO-2 Candidates}\\
\hline
Catalog Name&Hip. Number  &$\alpha(ICRS)$  & $\ $ $\delta(ICRS)$  & $\mu_{\alpha}\ \ \ \ $  & $\mu_{\delta}\ \ \ \ $  & V\ \ \        &$K_{s}\ $  & d$_{kin}$ & d$_{\pi}$ & X-ray CR& HR1& HR2 &EW(H$\alpha$)  &LNM\\[-0.07 in]
                           &                   &  (2000.0)             &  (2000.0)$\ $             & mas yr$^{-1}$    & mas yr$^{-1}$    & mag   & mag       & pc\ \               & pc                & k\ s$^{-1}\ \ \ $ &         &          & \AA\ \ \ \ \ \               &BPMG?\\
\hline  
TYC 5847-682-1  &                            & 10.532500   & -16.967806  & 67.7$\pm$3.2 & -32.0$\pm$3.4 & 11.18  & 8.02  & 55.6  & $\cdots$  & $\cdots$                & $\cdots$           & $\cdots$        & n.d.             &no\\[-0.08 in]
TYC 3681-1064-1  &                              & 17.726208  & 58.370472   & 78.5$\pm$3.6 & -48.6$\pm$3.9  & 12.65  & 8.22  & 45.5  & $\cdots$  & 99$\pm$19    &  -0.37$\pm$0.18  & 0.63$\pm$0.27  & $\cdots$      & table 4\\[-0.08 in]
TYC 1208-468-1  &                           & 24.413958  & 18.592528   & 69.9$\pm$1.5 & -47.5$\pm$1.5  & 10.59  & 6.72  & 52.6  & $\cdots$  & 753$\pm$43    &  -0.13$\pm$0.05  & 0.15$\pm$0.08 & 1.9       & table 4\\[-0.08 in]
TYC 1766-1431-1  & HIP 11152     & 35.860708 & 22.735528   &  95.6$\pm$1.9 &  -109.1$\pm$2.1 &11.36  & 7.35  & 29.4  & 30.7$\pm$2.5   & 453$\pm$95    &  0.04$\pm$0.20  & 0.22$\pm$0.29 & 2.1       & table 4\\[-0.08 in]
TYC 7558-655-1  &                           & 37.634792  & -43.706444   & 80.5$\pm$1.6 &  -14.9$\pm$1.6  &10.42  & 7.23  & 35.7  & $\cdots$   & 175$\pm$24    &  -0.05$\pm$0.13  & 0.19$\pm$0.18 & n.d.       & table 4\\[-0.08 in]
TYC 1926-794-1 &                           & 120.431792 & 23.707806   & -104.4$\pm$1.3 &  -116.9$\pm$1.3  &10.99& 6.80  & 25.0  & $\cdots$   & $\cdots$    &  $\cdots$  & $\cdots$ & n.d.     & no\\[-0.08 in]
TYC 8227-1561-1 &                        & 174.559125  & -51.032444  & -76.6$\pm$2.3 &  -42.3$\pm$2.1  &11.31 &  7.84 &  45.5  & $\cdots$   & $\cdots$    &  $\cdots$  & $\cdots$ & n.d.     & no\\[-0.08 in]
TYC 5643-155-1 &                        & 255.411000    &-7.727250  & -33.9$\pm$2.0 & -72.6$\pm$2.1   &10.50     &  6.14 &  37.0  & $\cdots$   & $\cdots$    &  $\cdots$  & $\cdots$ & n.d.     & no\\[-0.08 in]
TYC 5092-356-1 &                        & 263.694750	   & -6.902806  & -18.0$\pm$4.6 &  -83.7$\pm$4.8    &11.34     &  7.45 &  32.3  & $\cdots$   & $\cdots$    &  $\cdots$  & $\cdots$ & n.d.     & no\\[-0.08 in]
TYC 9114-1267-1  &                        & 320.369125  & -66.918194  & 96.9$\pm$2.8 &  -96.6$\pm$2.6  &10.60     &  7.01 &  31.3  & $\cdots$   & $\cdots$    &  $\cdots$  & $\cdots$ & n.d.     & no\\[-0.08 in]
TYC 1158-1185-1  &                           & 340.703625  & 13.514694    & 73.7$\pm$1.9 &  -41.2$\pm$2.0  &11.51  & 7.96  & 47.6  & $\cdots$   & 203$\pm$24    &  0.06$\pm$0.11  & 0.03$\pm$0.16  & 1.5      & table 4\\
\hline
\multicolumn{15}{c}{SUPERBLINK Candidates$^{a}$}\\
\hline
PM I01192-0126 &                         & 19.814917	        &-1.445306    & 40.0   & -19.0     & 12.47   & 8.81  &  90.9  & $\cdots$   & $\cdots$    &  $\cdots$  & $\cdots$ & n.d.     & no\\[-0.08 in]
PM I02302+0300&                         & 37.550000	        &3.015611    & 36.0   & -26.0    & 13.41   & 9.42  &  83.3  & $\cdots$   & $\cdots$    &  $\cdots$  & $\cdots$ & n.d.     & no\\[-0.08 in]
PM I04146+1829&                         & 63.674208		&18.487528    & 23.0   & -40.0     & 13.18   & 9.36  &  76.9  & $\cdots$   & $\cdots$    &  $\cdots$  & $\cdots$ & n.d.     & no\\[-0.08 in]
PM I04439+3723W &                        & 70.986958       &37.384250    & 18.0   & -52.0     & 13.41  & 8.80  &  76.9  & $\cdots$   & 95$\pm$17    &  -0.59$\pm$0.13  & 0.79$\pm$0.35  & $\cdots$     & table 4\\[-0.08 in]
PM I05231+2049&                         & 80.784083	         &20.821222    & 13.0   & -46.0     & 12.89   & 8.59  &  76.9  & $\cdots$   & $\cdots$    &  $\cdots$  & $\cdots$ & n.d.     & no\\[-0.08 in]
PM I06097+1607&                         & 92.443625		&16.118583    & -3.0   & -44.0     & 12.05   & 8.69  &  76.9  & $\cdots$   & $\cdots$    &  $\cdots$  & $\cdots$ & n.d.     & no\\[-0.08 in]
PM I06562+1813&                         & 104.051083        &18.223333    & -7.0   & -47.0     & 14.00   & 9.31  &  71.4  & $\cdots$   & $\cdots$    &  $\cdots$  & $\cdots$ & n.d.     & no\\[-0.08 in]
PM I23213+2448&                         & 350.346750        & 24.803167    & 45.0   & -32.0     & 12.90   & 8.65 &  76.9  & $\cdots$   & $\cdots$    &  $\cdots$  & $\cdots$ & n.d.     & no\\
\hline
\multicolumn{15}{c}{AB Doradus Interlopers}\\
\hline
TYC 8995-225-1&  HIP 31878    & 99.958557    & -61.478378   & -25.7$\pm$2.0   &  71.3$\pm$2.5   & 9.69   & 6.50 &  29.4  &  22.4$\pm$0.5 & 143$\pm$14    &  -0.05$\pm$0.09  & -0.13$\pm$0.13  & n.d.     & no\\
\hline
\multicolumn{15}{l}{\scriptsize{$^{a}$ Error in SUPERBLINK proper motions is $\pm$8 mas yr$^{-1}$}}\\
\end{tabular}
\end{sidewaystable}

\clearpage

\begin{sidewaystable}
\scriptsize
\hspace{-2cm}
\begin{tabular}{llrrrrrrrrrrrrr}
\multicolumn{15}{c}{\textbf{Table 2}}\\
\multicolumn{15}{c}{\textbf{Candidate AB Doradus Moving Group Members}}\\
\hline
\multicolumn{15}{c}{TYCHO-2 Candidates}\\
\hline
Catalog Name&Hip. Number  &$\alpha(ICRS)$  & $\ $ $\delta(ICRS)$  & $\mu_{\alpha}$\ \ \ \   & $\mu_{\delta}$\ \ \ \   & V\ \ \         &$K_{s}$\   & d$_{kin}$ & d$_{\pi}$ & X-ray CR & HR1& HR2 & EW(H$\alpha$)   &LNM\\[-0.07 in]
                           &                   &  (2000.0)             &  (2000.0)$\ $             & mas yr$^{-1}$    & mas yr$^{-1}$    & mag   & mag       & pc\ \               & pc               & k\ s$^{-1}$\ \ \ &         &          & \AA\ \ \ \ \ \           &ABDMG?\\
\hline  
TYC 1741-2117-1  &                         & 8.534750	    & 25.397361  & 93.5$\pm$2.1 &  -97.3$\pm$2.1  & 11.18  & 7.66  & 47.6 & $\cdots$  & 174$\pm$25    &  -0.21$\pm$0.14  & 0.00$\pm$0.22  & 2.1      & table 4\\[-0.08 in]
TYC 8474-1184-1  &                         & 20.325167   &-54.573417   & 85.1$\pm$3.7 &  -43.8$\pm$3.4  & 11.28  & 7.81  & 47.6  & $\cdots$  & $\cdots$   & $\cdots$    &  $\cdots$ & n.d.     & no\\[-0.08 in]
TYC 7002-2219-1  &                         & 20.518167   & -33.617556   & 107.8$\pm$3.8 &  -56.0$\pm$3.9  & 11.16  & 7.45  & 43.5  & $\cdots$  & 317$\pm$35    &  -0.23$\pm$0.10  & -0.14$\pm$0.17  &1.7     & table 4\\[-0.08 in]
TYC 6426-182-1  &                           & 22.512875   &-22.509639   & 115.4$\pm$2.7 &  -71.8$\pm$3.0 & 10.76  & 7.48  & 41.7  & $\cdots$  & $\cdots$   & $\cdots$    &  $\cdots$  &  n.d.     & no\\[-0.08 in]
TYC 1752-63-1  &                              & 24.346500  & 26.953667   & 119.1$\pm$2.1 &  -126.8$\pm$2.1  & 10.77  & 7.64  & 38.5  & $\cdots$  & 251$\pm$31    &  -0.07$\pm$0.11  & -0.12$\pm$0.17  & $\cdots$     & table 4\\[-0.08 in]
TYC 1252-798-1  &                           & 56.561708  &17.153028   & 52.2$\pm$2.5 &  -120.8$\pm$2.5  & 11.40  & 8.24  & 47.6  & $\cdots$  & $\cdots$   & $\cdots$    &  $\cdots$  &  n.d.     & no\\[-0.08 in]
TYC 1254-666-1  &                           & 61.567375  &17.120889     & 81.3$\pm$2.1 & -147.7$\pm$2.1 & 11.24  & 7.82  & 37.6  & $\cdots$  & $\cdots$   & $\cdots$    &  $\cdots$  &  n.d.     & no\\[-0.08 in]
TYC 90-406-1       &                           & 68.126708  &4.139889     & 43.1$\pm$1.9 &  -175.6$\pm$2.1  & 11.49  & 7.64  & 30.3  & $\cdots$  & $\cdots$   & $\cdots$    &  $\cdots$  &  n.d.     & no\\[-0.08 in]
TYC 5899-26-1  &                         & 73.101417	&-16.822278    & 118.9$\pm$4.5 &   -211.9$\pm$4.7  & 11.64  & 6.89  & 14.9  & $\cdots$  & 431$\pm$35    &  -0.22$\pm$0.07  & -0.02$\pm$0.12  & $\cdots$     & table 4\\[-0.08 in]
TYC 8513-952-3  & HIP 22738     & 73.379500	 & -55.860472   & 130.2$\pm$3.0  &  73.3$\pm$2.9  &11.00  & 6.34  & 11.6  & 11.2$\pm$0.4   & 1008$\pm$108    &  -0.20$\pm$0.09  & 0.12$\pm$0.15  & 7.7      & table 4\\[-0.08 in]
TYC 9388-888-1  &                           & 102.714125  &-79.963833  & -8.3$\pm$2.2  &   75.4$\pm$2.1   & 10.81 & 7.67  & 50.0  & $\cdots$  & $\cdots$   & $\cdots$    &  $\cdots$  &  n.d.     & no\\[-0.08 in]
TYC 9389-53-1  &                           & 117.695792  &-79.867778 & -48.0$\pm$2.9   &   57.7$\pm$2.8   & 8.66 & 7.81  & 50.0  & $\cdots$  & $\cdots$   & $\cdots$    &  $\cdots$  &  n.d.     & no\\[-0.08 in]
TYC 5417-1424-1  &                           & 120.776542	 &-9.686639 & -37.9$\pm$2.3  &  -99.7$\pm$2.3 & 10.93 &7.68  & 43.5  & $\cdots$  & $\cdots$   & $\cdots$    &  $\cdots$  &  n.d.     & no\\[-0.08 in]
TYC 812-1445-1  &                           & 136.105333	& 9.801639 & -31.3$\pm$3.7  &   -80.2$\pm$3.6   & 11.50 & 8.20  & 71.4  & $\cdots$  & $\cdots$   & $\cdots$    &  $\cdots$  &  n.d.     & no\\[-0.08 in]
TYC 1405-1574-1  &                           & 140.772583	& 19.157861 & -46.6$\pm$2.6  &  -77.9$\pm$2.7   & 12.00 & 8.89  & 71.4  & $\cdots$  & $\cdots$   & $\cdots$    &  $\cdots$  &  n.d.     & no\\[-0.08 in]
TYC 1406-795-1  &                           & 143.934583	& 18.597083 & -56.0$\pm$1.4  & -104.2$\pm$1.7  & 10.83 & 7.71  & 55.6  & $\cdots$  & $\cdots$   & $\cdots$    &  $\cdots$  &  n.d.     & no\\[-0.08 in]
TYC 6068-1130-1  &                           & 154.968042 & -18.991639 & -67.4$\pm$3.1  &  -46.2$\pm$3.1    & 11.39 & 8.21  & 66.7  & $\cdots$  & $\cdots$   & $\cdots$    &  $\cdots$  &  n.d.     & no\\[-0.08 in]
TYC 8966-1336-1  &                           & 163.369375 & -65.408250 &  -96.1$\pm$4.9   &   -24.0$\pm$4.7  & 12.31& 8.42  & 45.5  & $\cdots$  & $\cdots$   & $\cdots$    &  $\cdots$  &  n.d.     & no\\[-0.08 in]
TYC 5514-1135-1  &  HIP 55066      & 169.092708	& -14.693111 & -175.1$\pm$1.7  & -120.1$\pm$1.7   & 10.07 & 6.46 & 29.4  & 17.6$\pm$0.5  & $\cdots$   & $\cdots$    &  $\cdots$  &  n.d.     & no\\[-0.08 in]
TYC 4933-773-1  &  HIP 56367      & 173.320208	& -3.402000 & -120.4$\pm$2.0  &  -116.5$\pm$2.0    & 11.29 & 7.53 & 38.5  & 33.6$\pm$3.2  & $\cdots$   & $\cdots$    &  $\cdots$  &  n.d.     & no\\[-0.08 in]
TYC 8227-1561-1  &                           & 174.559125	& -51.032444 & -76.6$\pm$2.3  &  -42.3$\pm$2.1    & 11.31 & 7.84  & 58.8  & $\cdots$  & $\cdots$   & $\cdots$    &  $\cdots$  &  n.d.     & no\\[-0.08 in]
TYC 4943-192-1 &                             & 183.826875 &	-2.624278   &  -83.8$\pm$3.8   &  -118.5$\pm$4.0    &11.36  & 7.83  & 45.5  & $\cdots$  & 69$\pm$23   &  0.25$\pm$0.32  & -0.24$\pm$0.55  & 1.5       & table 4\\[-0.08 in]
TYC 6709-457-1  &                           & 193.643958 &	-29.855000 &  -53.1$\pm$4.5   & -69.6$\pm$4.2   & 11.64 & 8.32  & 71.4  & $\cdots$  & $\cdots$   & $\cdots$    &  $\cdots$  &  n.d.     & no\\[-0.08 in]
TYC 8307-34-1  & HIP 76107       & 233.153292 &	-52.355444 & -98.0$\pm$4.6 &  -158.6$\pm$4.3   & 11.17 & 7.60  & 34.5  &  30.5$\pm$5.0  & $\cdots$   & $\cdots$    &  $\cdots$  &  n.d.     & no\\[-0.08 in]
TYC 7337-167-1  &                           & 238.883250 &	-33.758389 & -45.9$\pm$2.5  &  -85.6$\pm$ 2.7   & 12.14 &8.83  & 66.7  & $\cdots$  & $\cdots$   & $\cdots$    &  $\cdots$  &  n.d.     & no\\[-0.08 in]
TYC 9052-2737-1  & HIP 84301       &258.529667 &	-60.243861 & -11.1$\pm$6.7  &   -126.4$\pm$6.0   & 10.62 & 7.38  & 50.0  &  76.5$\pm$52.7  & $\cdots$   & $\cdots$    &  $\cdots$  &  n.d.     & no\\[-0.08 in]
TYC 9053-228-1  &                           & 262.195750 &	-61.729611 & -7.4$\pm$4.8   & -73.7$\pm$4.3   & 12.30 & 8.72  & 83.3  & $\cdots$  & $\cdots$   & $\cdots$    &  $\cdots$  &  n.d.     & no\\[-0.08 in]
TYC 9063-2884-1  &                           & 272.068542 & -65.197250 &  -17.4$\pm$4.7 &  -109.4$\pm$4.4  & 12.06 & 8.28  & 55.6  & $\cdots$  & $\cdots$   & $\cdots$    &  $\cdots$  &  n.d.     & no\\[-0.08 in]
TYC 7392-1588-1  &                           & 273.117792 & -30.224389 & 19.1$\pm$3.4 &   -236.6$\pm$3.4   & 11.07 & 7.03 & 27.0  & $\cdots$  & $\cdots$   & $\cdots$    &  $\cdots$  &  n.d.     & no\\[-0.08 in]
TYC 8757-314-1  &                           & 277.550167 &	-58.273278 & -25.3$\pm$2.2  &  -446.7$\pm$ 2.3   & 9.84 & 5.96 & 14.3  & $\cdots$  & $\cdots$   & $\cdots$    &  $\cdots$  &  n.d.     & no\\[-0.08 in]
TYC 7407-664-1  &                           & 280.323500 &	-30.768444 & -7.5$\pm$4.6 &  -205.3$\pm$4.3  & 11.38 & 7.23 & 31.3  & $\cdots$  & $\cdots$   & $\cdots$    &  $\cdots$  &  n.d.     & no\\[-0.08 in]
TYC 5126-181-1  &                           & 283.052542 &	-7.225167 & 67.7$\pm$2.8 &  -250.7$\pm$3.1  & 11.32 & 6.82 & 20.8  & $\cdots$  & $\cdots$   & $\cdots$    &  $\cdots$  &  n.d.     & no\\[-0.08 in]
TYC 7921-2351-1  &                           & 286.045917 & -40.445250 & 21.6$\pm$1.7   &  -148.0$\pm$2.1  & 10.61 & 7.22 & 43.5  & $\cdots$  & $\cdots$   & $\cdots$    &  $\cdots$  &  n.d.     & no\\[-0.08 in]
TYC 7451-1555-1  &                           & 297.442042 & -35.872667 & 30.4$\pm$2.6  &   -65.5$\pm$2.8   & 12.16 & 9.00 & 90.9  & $\cdots$  & $\cdots$   & $\cdots$    &  $\cdots$  &  n.d.   & no\\[-0.08 in]
TYC 523-573-1  &                         & 309.977292 & 6.336861    & 89.5$\pm$2.2  &   -100.1$\pm$2.1  & 10.53  & 7.13  & 38.5  & $\cdots$  & 70$\pm$14   &  -0.4$\pm$0.18  & 0.13$\pm$0.38  & $\cdots$    & table 4\\[-0.08 in]
TYC 7465-297-1  &                           & 311.287958 &	-33.949639 & 52.3$\pm$3.1  &  -85.1$\pm$3.2   & 12.69 & 9.03 & 66.7  & $\cdots$  & $\cdots$   & $\cdots$    &  $\cdots$  &  n.d.    & no\\[-0.08 in]
TYC 6934-569-1  & HIP 102817       &312.449667 &	-28.468806 & 76.5$\pm$3.1 &  -153.0$\pm$3.0   & 10.83 & 7.41  & 38.5  &  44.2$\pm$4.7  & $\cdots$   & $\cdots$    &  $\cdots$  &  n.d.    & no\\[-0.08 in]
TYC 9331-189-1  & HIP 107705     & 327.272083 &	-72.101833  & 320.2$\pm$2.1 &   -302.5$\pm$2.0   &9.96  & 5.65  & 12.5  & 16.1$\pm$0.6   & 861$\pm$73    &  -0.17$\pm$0.08  & -0.13$\pm$0.12   & 2.3  & no\\[-0.08 in]
TYC 5809-128-1  & HIP 109528     & 332.800667 &	-12.861444  & 75.9$\pm$2.1  &  -99.1$\pm$2.1   &11.22  & 7.79  & 52.6 & 41.5$\pm$3.6  & $\cdots$   & $\cdots$    &  $\cdots$  &  n.d.    & no\\[-0.08 in]
TYC 8003-813-1  &  HIP 111078     & 337.540417	& -44.379444 & 113.7$\pm$2.3  &   -109.6$\pm$2.2   & 11.23 & 7.68 & 40.0  & 34.9$\pm$6.8  & $\cdots$   & $\cdots$    &  $\cdots$  &  n.d.    & no\\[-0.08 in]  
TYC 9337-2134-1  &                         & 341.785792 &  -69.345556  & 63.3$\pm$4.5  &  -59.9$\pm$4.3     &11.22  & 8.09  & 62.5  & $\cdots$   & 68$\pm$15    &  -0.23$\pm$0.19  & 0.24$\pm$0.30    & 1.3  & table 4\\
\hline
\multicolumn{15}{c}{$\beta$ Pictoris Interlopers}\\
\hline
TYC 9172-690-1&  HIP 29964    & 94.617625	    & -72.045028  & -8.5   &  75.7   & 10.00   & 6.81 &  38.5  &  38.6$\pm$1.5 & 1030$\pm$13    &  -0.07$\pm$0.02  & 0.03$\pm$0.04    & 1.9  & no\\
\hline
\end{tabular}
\end{sidewaystable}
\clearpage

\begin{sidewaystable}[!h]
\begin{center}
\scriptsize
\begin{tabular}{llrrrrrr}
\multicolumn{8}{c}{\textbf{Table 3}}\\
\multicolumn{8}{c}{\textbf{RV Data of Selected Known AB Doradus Moving Group Members}}\\
\hline\hline
Catalog Name   &Alt. Name    &$\alpha(ICRS)$    &$\delta(ICRS)$    &RV$_{p\ \ \ }$     &RV$_{m\ \ \ \ }$    & $v$sin$i$\ \ \  &SpTy\\                         
    &                     &  (2000.0)$\ $                &  (2000.0)$\ $                & km\ s$^{-1}$     & km\ s$^{-1}\ \ $    & km\ s$^{-1}$   &   \\
\hline
\hline\hline

TYC 1758-1794-1   & HIP 10272    & 33.063895    & 23.958569    & -0.2    & -0.7$\pm$0.5    & 6$\pm$2 &K1III  \\ 
TYC 2845-2242-1    & HIP 12635    & 40.587083    &  38.622823    & -4.6    & -5.6$\pm$0.7    & 2$\pm$2& K4V  \\
TYC 2845-2243-1    & HIP 12638    & 40.588569     & 38.618918    & -4.6    & -5.7$\pm$0.5    & 2$\pm$2 & G5V  \\
TYC 1231-929-1    & HIP 14807    & 47.801248    & 22.423307    & +4.6    & +3.1$\pm$0.6    & 6$\pm$2 & K6V  \\
TYC 1231-536-1    & HIP 14809    & 47.807544    &  22.416156    & +4.7    & +4.4$\pm$0.7    & 2$\pm$2 & G5V \\
TYC 3312-2141-1    & V577 Per    & 53.305983     & 46.257787    & -5.4    & -8.4$\pm$0.8    & 2$\pm$2 & K2V  \\ 
TYC 73-264-1     &  HIP 19183     & 61.672974    & 1.684138    & +17.1    & +17.6$\pm$2.5    & 6$\pm$2 & F5V  \\
TYC 2738-1390-1    & GJ 856A    & 335.870579    &  32.459942    & -21.0    & -20.0$\pm$1.2    & 6$\pm$2 & M3V  \\
TYC 3225-2271-1    & HIP 115162    & 349.914600    &  42.252882    &    -19.5    & -22.1$\pm$0.9    & 2$\pm$2 & G0V  \\
\hline
\end{tabular}
\end{center}
\end{sidewaystable}

\clearpage

\begin{sidewaystable}
\begin{center}
\scriptsize
\begin{tabular}{llrrrrrrlr}
\multicolumn{10}{c}{\textbf{Table 4}}\\
\multicolumn{10}{c}{\textbf{RV Data of PYCs}}\\
\hline\hline
Catalog Name  &Alt. Name   &$\alpha(ICRS)$   &$\delta(ICRS)$                   &RV$_{p\ \ }^{{\ a}}$\ \     &RV$_{m\ \ \ \ }$   & $v$sin$i$\ \ \  &  d$_{kin}$ &SpTy$^{\ }$   &     LNM?\\
                            &                    &  (2000.0)$\ $               &  (2000.0)$\ $               & km\ s$^{-1}$            & km\ s$^{-1}\ \ $            & km\ s$^{-1}$ &   pc\ \                       &                            \\
\hline
\\
\multicolumn{10}{c}{\textbf{BPMG PYCs}}\\
\hline\hline
TYC 1208-468-1   & BD+17 232   & 24.413958   &  18.592528   & +3.9\ \ \    &  +3.2$\pm$1.0$^{b}$  & $\cdots$$^{\ }$  & 52.6  & K3Ve$^{b}$    & YES$^{\ }$\\
TYC 1208-468-2   & BD+17 232B   & $\cdots$    &  $\cdots$    &  +3.9\ \ \  & $\cdots$$^{\ }$ & $\cdots$$^{\ }$  & 52.6  &  $\cdots$$^{\ }$   & YES$^{\ }$\\
TYC 1766-1431-1  & HIP 11152   & 35.860708   & 22.735528   & +6.6\ \ \  & +10.4$\pm$2.0$^{\ }$ & 6$\pm$2 & 29.4  &    M3Ve$^{\ }$   & YES$^{\ }$ \\ 
TYC 7558-655-1   & CD-44 753  &  37.634792   & -43.706444   & +15.8\ \ \    & +16.3$\pm$1.1$^{c}$  &  $\cdots$$^{\ }$    &  35.7 &   K5Ve$^{c}$   & YES$^{\ }$\\
PM I04439+3723W &  $\cdots$   & 70.986958   &37.384250 & +6.0\ \ \ & +6.2$\pm$2.0$^{\ }$ & 8$\pm$2  & 76.9  &  M3Ve$^{\ }$   &  YES$^{\ }$\\
PM I04439+3723E  &  $\cdots$   & 70.989624      & 37.384178   &  +6.0\ \ \  & $\cdots$  & $\cdots$$^{\ }$  & 76.9   & $\cdots$          &  YES$^{\ }$\\
TYC 1158-1185-1N  & RBS 1885   & 340.703625   &  13.514694   & -7.4\ \ \    & -14.9$\pm$1.5$^{\ }$ &  2$\pm$2 & 47.6     & K5Ve$^{\ }$   & NO$^{\ }$\\
TYC 1158-1185-1S  & $\cdots$   & $\cdots$    & $\cdots$     & -7.4\ \ \    & -14.8$\pm$1.5$^{\ }$ & 2$\pm$2  & 47.6     & K5Ve$^{\ }$   & NO$^{\ }$\\
\hline
\\
\multicolumn{10}{c}{\textbf{ABDMG PYCs}}\\
\hline\hline
TYC 1741-2117-1N    & GSC 01741-02117   & 8.534750   & 25.397361   & -8.7\ \ \   & -12.4$\pm$2.0$^{\ }$ & 10$\pm$2 & 47.6    & K7Ve$^{\ }$  & YES$^{\ }$\\
TYC 1741-2117-1S    & $\cdots$  & $\cdots$   & $\cdots$   & -8.7\ \ \   & -10.8$\pm$2.0$^{\ }$ & 10$\pm$2 & 47.6    & K7Ve$^{\ }$  & YES$^{\ }$\\
TYC 7002-2219-1  & $\cdots$  & 20.518167   & -33.617556   & +18.3\ \ \   & +3.0$\pm$1.4$^{\ }$& 6$\pm$2  & 43.5    & K7Ve$^{\ }$  & NO$^{\ }$\\
TYC 1752-63-1  & StKM 1-174 & 24.346500  & 26.953667  &    -4.1\ \ \ & -5.9$\pm$3.0$^{\ }$  & 10$\pm$2  & 38.5 & K5Ve$^{\ }$ & YES$^{\ }$\\
TYC 5899-26-1   & $\cdots$    & 73.101417  & -16.822278   & +26.3\ \ \  & +26.7$\pm$1.5$^{\ }$& 0$\pm$2  &  14.9   & M3Ve$^{\ }$  &  YES$^{d}$\\
TYC 8513-952-3   & HIP 22738B   & 73.377083 & -55.858889  &  +30.4\ \ \ & +30.0$^{c}$ & $\cdots$$^{\ }$ &11.6 & M3Ve$^{c}$   &  YES$^{d}$ \\
TYC 8513-952-1   & HIP 22738A   & 73.379500  & -55.860472  & +30.4\ \ \ & +29.0$^{c}$    & $\cdots$$^{\ }$  &11.6  & M3Ve$^{c}$   &  YES$^{d}$ \\
TYC 4943-192-1  & RAVE J121518.4-023728   & 183.826875  & -2.624278  & +0.3\ \ \  & +1.4$\pm$4.6$^{e}$ & $\cdots$$^{\ }$  & 45.5    & M0Ve$^{f}$   &  YES$^{\ }$ \\
TYC 523-573-1   & BD+05 4576   & 309.977292  & 6.336861   & -19.6\ \ \    & -20.6$\pm$1.5$^{\ }$ & 6$\pm$2  & 38.5   & K7Ve$^{\ }$  & YES$^{\ }$\\
TYC 9337-2134-1    & $\cdots$   & 341.785792  & -69.345556   & +18.6\ \ \    & -2.2$^{c}$ & $\cdots$$^{\ }$ & 62.5    & K6Ve$^{c}$  & NO$^{\ }$\\
\hline
\multicolumn{10}{l}{\scriptsize{$^{a}$ Typical error in RV$_{p}$ is $\pm$2.0 km s$^{-1}$ (the typical UVW velocity spread in the NYMGs)}}\\
\multicolumn{10}{l}{\scriptsize{$^{b}$ Jeffries et al., 1995. $^c$ Torres et al., 2006. $^{d}$ Reconfirmation of Torres et al., 2008. }}\\
\multicolumn{10}{l}{\scriptsize{$^{e}$ Zwitter et al., 2008. $^{f}$ SpTy classified by visual inspection of spectra.}}
\end{tabular}
\end{center}
\end{sidewaystable}

\clearpage

\begin{figure}
\plotone{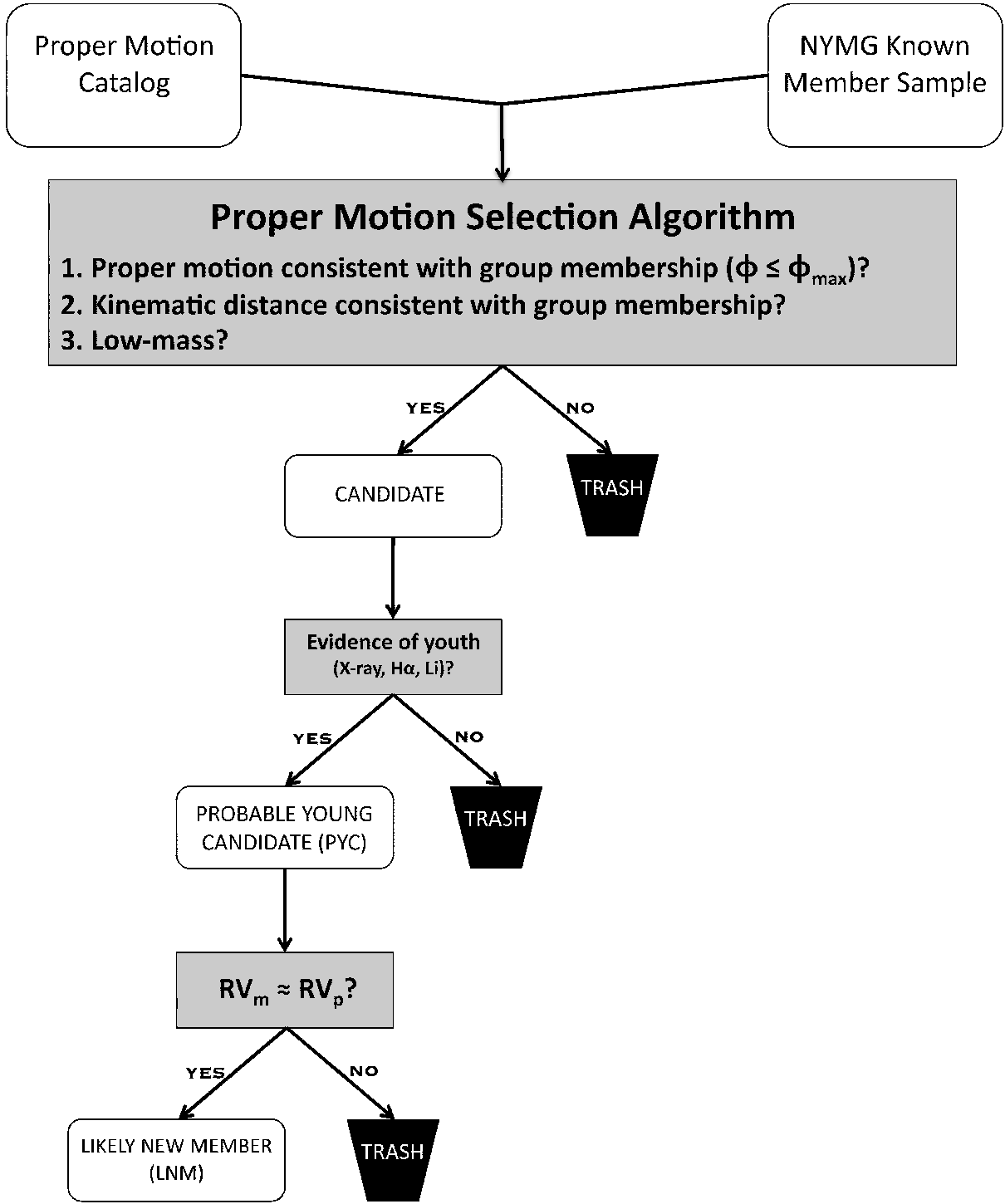}
\epsscale{0.85}
\caption{Steps in the identification of likely new members starting from proper motion and photometric data.}
\label{flowchart_plot}
\end{figure}

\clearpage

\begin{figure}
\epsscale{0.55}
\plotone{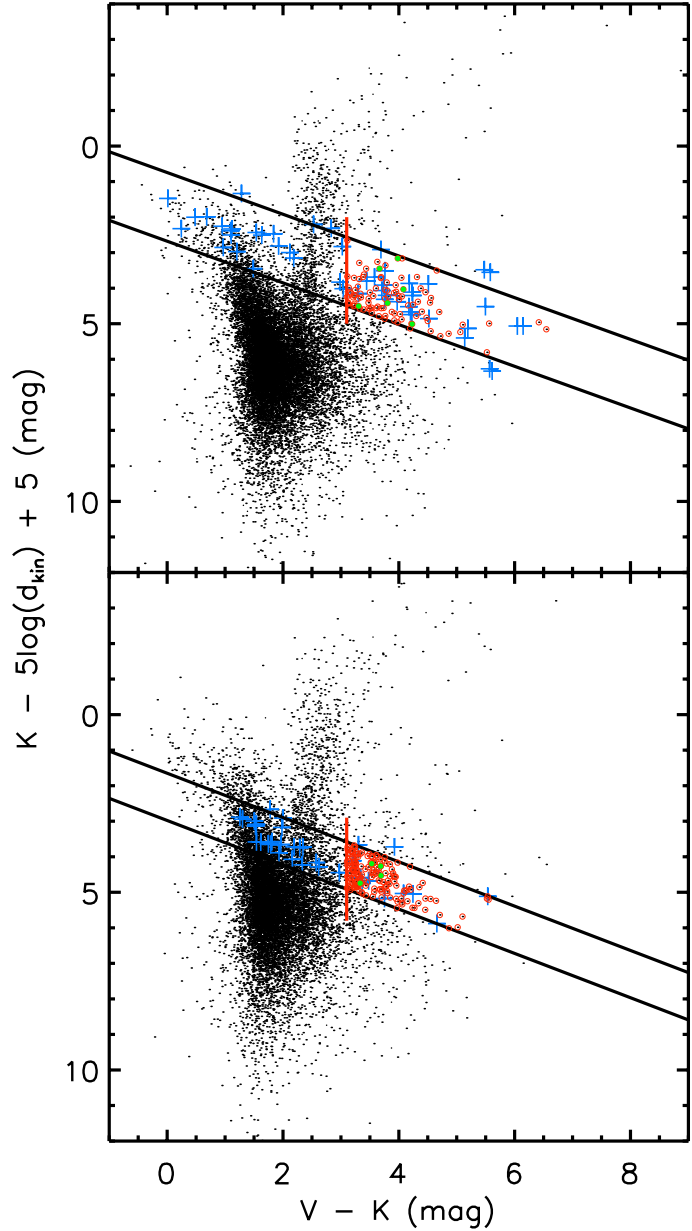}
\caption{\scriptsize{Top:  Color-magnitude diagram of BPMG TYCHO-2 candidates with known BPMG members (blue crosses).  Small dots represent the 16,031 TYCHO-2 stars in the $\mu$ $\ge$ 70 mas yr$^{-1}$ sample having proper motions consistent with BPMG group membership ($\phi$ $\le$ $\phi_{max}$).  Open red circles represent the 86 low-mass candidates (V-K $\ge$ 3.1, red vertical line) identified in the search and green filled circles highlight systems identified as LNMs in this project (this work and LS09).  The width of the BPMG cluster sequence locus (black parallel lines) is determined by scaling the uncertainty of a least-squares linear fit to the sequence until the width is consistent with that of LS09.  Bottom:  Color-magnitude diagram of ABDMG TYCHO-2 candidates with known ABDMG cluster sequence showing the 13,812 TYCHO-2 stars in the $\mu$ $\ge$ 70 mas yr$^{-1}$ sample with proper motions consistent with ABDMG membership, symbol designations are the same as those for the BPMG.  One hundred eighty low-mass candidates are targeted for follow up observations, with five LNMs being identified thus far (two are components in a visual binary).  The width of the ABDMG cluster sequence locus is defined by twice the uncertainty of a least-squares linear fit to the sequence.  Known members of the ABDMG are fainter on average when compared to the BPMG, which leads to more main sequence contamination and a larger candidate sample.} }
\label{abdor_plot}
\end{figure}

\clearpage

\begin{figure}
\epsscale{0.90}
\plotone{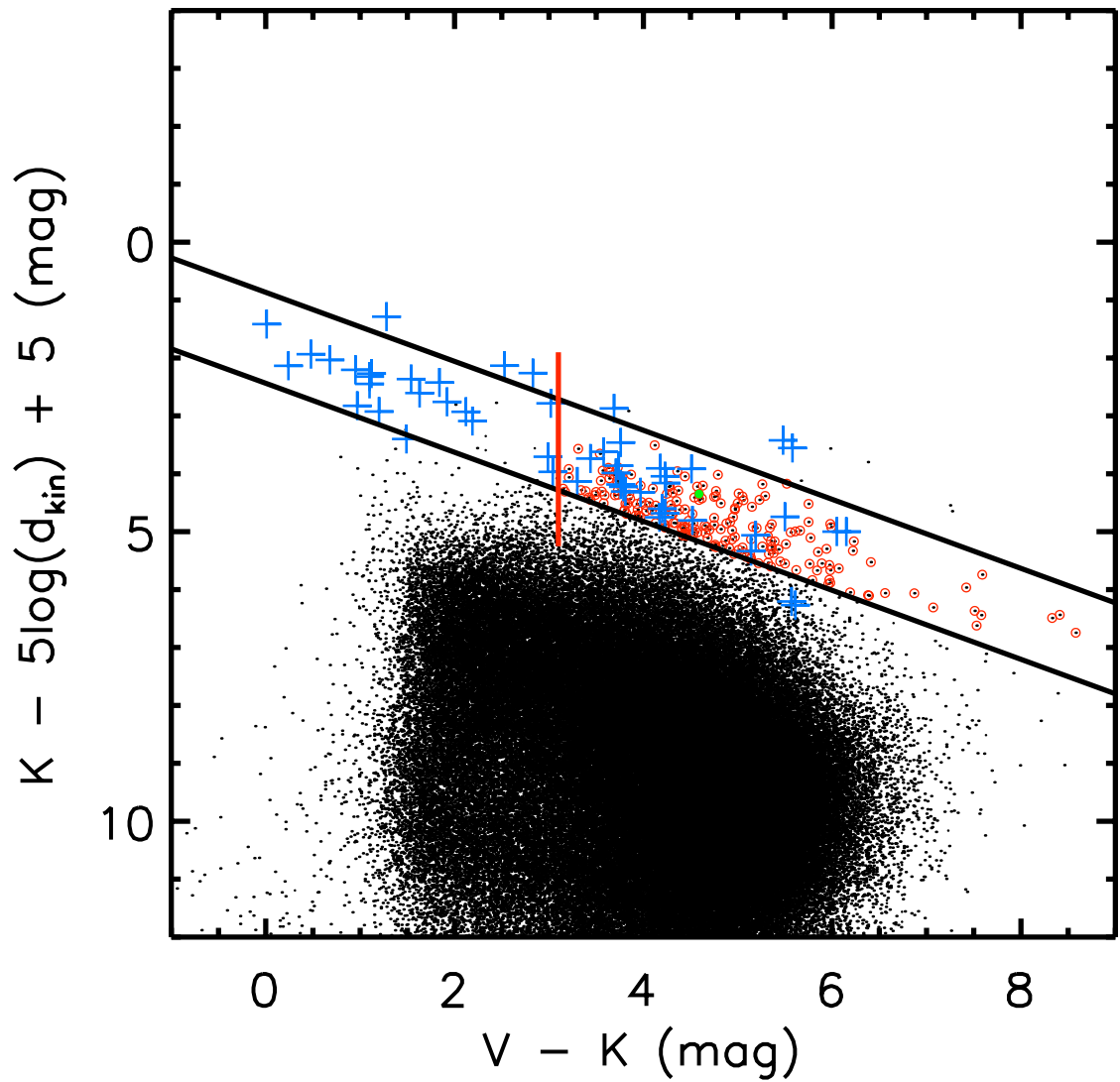}
\caption{Color-magnitude diagram of BPMG SUPERBLINK candidates with known BPMG members, symbol designations are the same as Fig. 2.  One hundred ninety-five low-mass candidates are identified from over 150,000 stars that have proper motion consistent with the BPMG.  Thus far we have identified two LNMs in one binary system from the sample of candidates.  The width of the BPMG cluster sequence locus is defined in the same way as the ABDMG locus in Fig. 2.  The SUPERBLINK catalog allows access to lower mass (larger V - K) candidates than the TYCHO-2 catalog. }
\label{bpic_plot}
\end{figure}

\clearpage

\begin{figure}
\epsscale{0.70}
\plotone{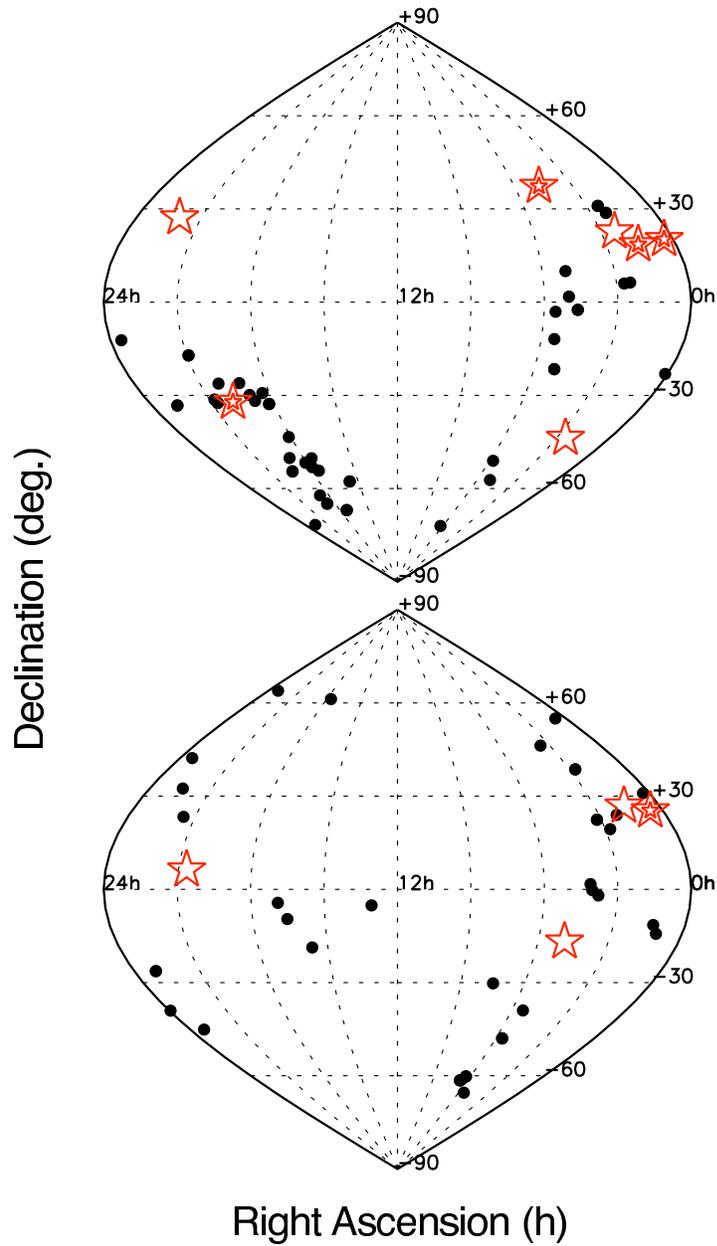}
\caption{Top:  Positions of known (filled circles) and likely new (red stars) $\beta$ Pictoris moving group members on the sky.  Bottom:  Positions of known and likely new AB Doradus moving group members on the sky.  LNMs represented with double star symbols are visual binaries.  Most LNMs identified thus far in our program are in the north.  We have doubled the number of likely $\beta$ Pictoris members in the northern hemisphere. }
\label{pos_plot}
\end{figure}

\clearpage

\begin{figure}
\epsscale{0.50}
\plotone{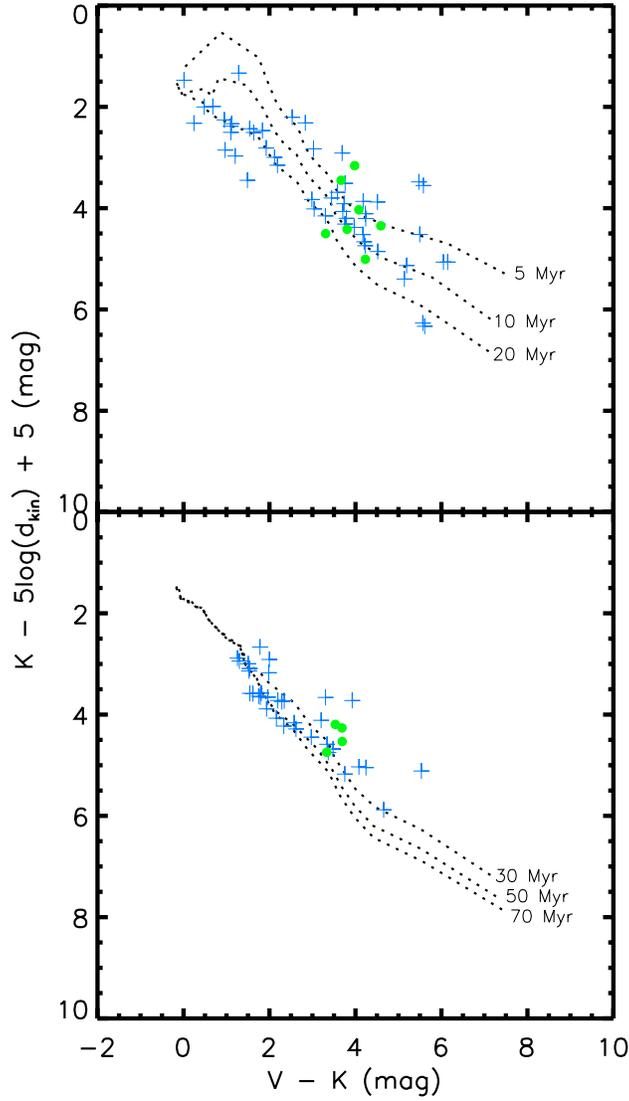}
\caption{Top:  Color-magnitude diagram of the BPMG showing known members (blue crosses) and LNM systems (green circles) with PMS isochrones of S2000 (dotted lines).  The CMD was generated using d$_{kin}$ to calculate M$_{K}$.  Isochrones illustrate the uncertainty in the age of the group with $\sim$10 Myr being the accepted value.  Bottom:  Color-magnitude diagram of the ABDMG using same symbol designations as top figure.  The ABDMG is older than the BPMG as illustrated by the isochrones.  All LNMs are consistent with group cluster sequences.}
\label{pos_plot}
\end{figure}

\clearpage

\end{document}